# Study of a MA-loaded cavity for compact synchrotron

SHI Hua (施华)[1;1)]    Yoshiro Irie (入江吉郎)[2]

[1] Key Laboratory of Particle Acceleration Physics and Technology，Institute of High Energy Physics, Chinese Academy of Sciences, Beijing 100049, China

[2] High Energy Accelerator Research Organization, 1-1 Oho, Tsukuba, Ibaraki 305-0801, Japan

**Abstract:** The untuned magnetic alloy (MA) loaded cavity will be adopted in the recent proposed projects based on compact proton and heavy-ion synchrotron in China. The MA loaded cavity has been operated successfully in many accelerators, but there is no construction and operation experience in China. The small-size MA cores with outer diameter >100 mm have been supplied by Chinese firms, and the properties of cores are tested and show good consistency with Hitachi material. Based on the MA cores testing results, the schematic design of a cavity is conducted, which could obtain 1 kV gap voltage with less than 1.5 kW power dissipation in the frequency range of 0.5~7 MHz. The analytic and simulation results were compared on the resonant frequency, shunt impedance and $Q$ value.

## 1  Introduction

Magnetic alloy (MA) loaded cavities have widely been used in proton and heavy-ion synchrotrons [1~7] due to the wide operating frequency range, untuned RF system and available high accelerating gradient. There are mainly three types of MA core used in RF cavity, and they are Fe-based amorphous metal [1], Co-based amorphous metal [2] and Nanocrystalline Fe-based soft magnetic material [3~7]. However, the Nanocrystalline Fe-based alloy, named as FINEMET in Hitachi, is adopted by most of the accelerators because of the good performance comparable to Fe-based amorphous metal and low cost comparable to Co-based amorphous metal. So the MA core with similar quality of FINEMET is required for developing the MA loaded cavity in China.

The study of constructing Nanocrystalline Fe-based soft magnetic material for accelerator is in the first stage in China, and the Chinese firms can provide only the small-size cores, the outer diameter of which is about 100 mm. Because the core property is the key factor for the performance of the RF system, several cores with different ribbon material and different annealing methods are tested and compared with Hitachi material. Based on the Chinese MA cores testing results, the schematic design example of low energy accelerating cavity with wide frequency range is conducted using analytic formula and the 3-D microwave simulation software CST.

## 2  Magnetic alloy cores

Comparing with soft ferrite material, the advantages of the MA-loaded RF cavities are summarized as follows.

1) Wide operating frequency band: the initial permeability of MA is very high, and the $\mu_s'$ (the real part of complex permeability) decreases with the frequency, so $\mu_s'$ is enough to satisfy the very wide frequency band.

2) High and stable shunt impedance: the parallel complex permeability (suffix p) is adopted to estimate the shunt impedance [8], and the relation between $\mu_s'$ and $\mu_p'$ is

$$\mu_p' = \mu_s'\left(1 + 1/Q_{ma}^2\right), \quad (1)$$

and $Q$-value of MA is

$$Q_{ma} = \mu_s'/\mu_s'', \quad (2)$$

where $\mu_s''$ is the imaginary part of complex permeability. The shunt impedance is

$$R_{ma} = n\mu_0 t \ln\left(\frac{r_o}{r_i}\right)\left(\mu_p' Q_{ma} f\right), \quad (3)$$

where $t$, $r_o$ and $r_i$ are the thickness, the outer and inner radii of the MA cores, respectively, and $\mu_0$ the vacuum permeability, $n$ the number of MA cores, and $f$ the paral-

---

[1)]Email: shih@ihep.ac.cn

lel circuit frequency. The $\mu_p'Q_{ma}f$-product of MA cores is high > $3\times10^9$ (Hz), so the impedance of MA-loaded cavity can become high enough. And also, the $\mu_p'$ and $Q_{ma}$ properties of the MA material are stable under varying RF magnetic field density ($B_{rf}$) and temperature.

3) Bias current source is not required and this makes RF system simple: because the MA cores have a low-$Q$ value of ~0.5 and high impedance, they are well suited to a tuning-free wide band cavity and it's unnecessary to have a bias current tuning system. The complicated bias winding in the cavity is dispensable and the whole RF system becomes simple and compact.

Chinese firms worked out several small MA cores and the specifications list in Table 1. The name of three kinds of cores in column 1 is given according to the name of materials in the column 5. In column 5, 1k107 is the Chinese 1k107 25 μm-thickness ribbon and FT-3 is the Hitachi's FT-3 18 μm-thickness ribbon. Column 6 means applying a magnetic field $H_v$ vertically or not to the core plane during annealing, and only 1k107-1 core did not apply $H_v$ during annealing. We tested magnetic properties of these cores with Network Analyzer (Agilent 4395A). Before the MA core test, we did the open, short and 50 Ω calibration. Then one MA core was wound with a short single-turn loop coil, and the impedance as a function of frequency from 0.5 to 20 MHz could be measured. Based on the measurement results, the $\mu_s'$, $\mu_s''$, $Q_{ma}$ and other parameters can be calculated.

Table 1. Specifications of the Chinese MA cores

| Name | $2r_o$ (mm) | $2r_i$ (mm) | $t$ (mm) | Material | Annealing |
|---|---|---|---|---|---|
| 1k107-1 | 100 | 50 | 25 | 1k107 (25 μm) |  |
| 1k107-2 | 130 | 60 | 25 | 1k107 (25 μm) | $H_v$ |
| FT-3 | 130 | 60 | 25 | FT-3 (18 μm) | $H_v$ |

Fig. 1 (a) and (b) show the frequency variation with permeability of 1k107-1 and 1k107-2 in 0.5~20 MHz frequency band. The ribbons of both cores are same, but 1k107-2 was annealed by vertical magnetic field which increases the complex permeability by a factor of 2~3.

Comparing with Hitachi cores, 1k107-1 properties are similar as FT3M-1 cores (shown in Fig. 2), which were made by Hitachi for JHF project [9] in the late 1990s. After improving the manufacture technology, as shown in Fig. 3, the $\mu_p'Q_{ma}f$-product of new Hitachi cores FT3M-2 (-■-) for J-PARC project [10] are about 40% higher than that of 1k107-1 (-▲-). Fig. 3 also shows 1k107-2 (-▼-) and FT-3 (-♦-), and we can see that properties of these two cores are similar as Hitachi FT3L cores (-□-), which were developed recently [10] with 13 μm-thickness ribbon and applying $H_v$ during annealing. But for producing 500 mm outer diameter cores, winding, annealing, coating and other processes would affect the magnetic properties of the cores, each one of which can affect the RF characteristics of the cores.

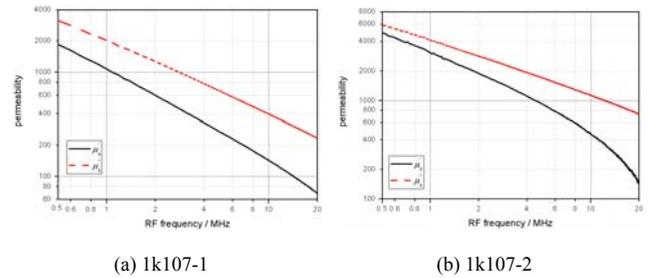

(a) 1k107-1   (b) 1k107-2

Fig. 1. Permeability versus frequency.

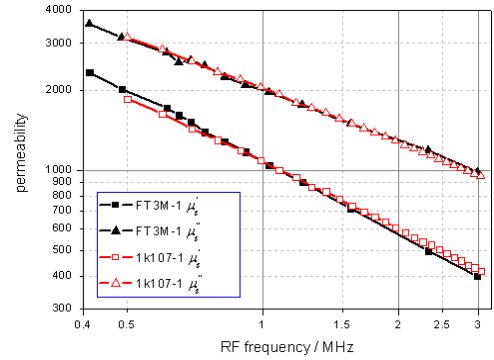

Fig. 2. Permeability comparison between 1k107-1 and FT3M-1.

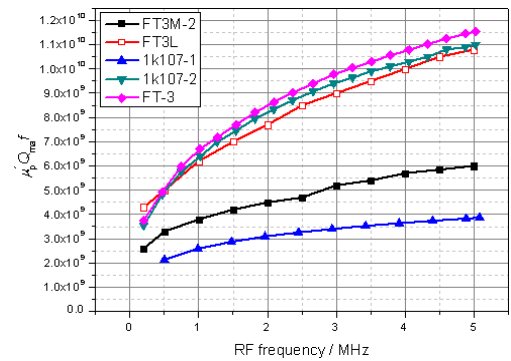

Fig. 3. $\mu_p'Q_{ma}f$-factor comparison between Chinese and Hitachi's cores.

In conclusion, for small MA cores, the ribbons and manufacturing processes of Chinese firms are similar as Hitachi, but it's necessary to further investigate the fabrication technology of large cores. Because very few Chinese firms have annealing oven with vertical magnetic field for large cores, we'll design the RF cavity using measurement data of the 1k107-1 in this report.

## 3 RF cavity

### 3.1 Design of RF cavity

We propose to adopt coaxial resonator loaded with several magnetic cores, which can be characterized as LC lumped cavity. Fig. 4 (a) shows the equivalent circuit, where $U_g$ is the gap voltage, $C$ the equivalent capacitance of the accelerating gap, $L_s$ the total inductance including the loaded magnetic cores and the empty cavity inductance, and $R_s$ the resistance to represent the total power loss in the magnetic cores.

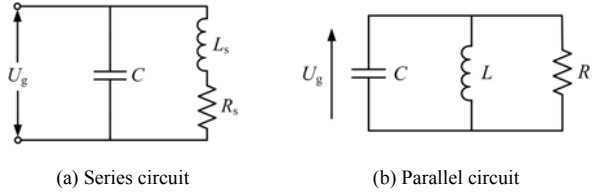

(a) Series circuit    (b) Parallel circuit

Fig. 4. Equivalent circuit of magnetic cores-loaded coaxial resonator.

$C$ is given by

$$C = \varepsilon_0 \varepsilon_r \frac{\pi \left( r_{gapo}^2 - r_{gapi}^2 \right)}{d_{gap}}, \quad (4)$$

where $\varepsilon_0$ is the vacuum permittivity, $\varepsilon_r$ the ceramic gap permittivity, $d_{gap}$, $r_{gapo}$, and $r_{gapi}$ the gap length, the radii of the outer and inner ceramic gap, respectively. $L_s$ is calculated by

$$L_s = L_{ma} + L_c, \quad (5)$$

where $L_{ma}$ is the inductance of loaded magnetic cores, and $L_c$ that of the empty cavity. They are obtained as

$$L_{ma} = n \frac{\mu_0}{2\pi} \mu_s' t \ln \frac{r_o}{r_i}, \quad (6)$$

$$L_c = \frac{\mu_0}{2\pi} \left( l_{cav} \ln \frac{r_{cav}}{r_{duct}} - nt \ln \frac{r_o}{r_i} \right). \quad (7)$$

$R_s$ is calculated by

$$R_s = n f \mu_0 \mu_s'' t \ln \frac{r_o}{r_i}, \quad (8)$$

where $l_{cav}$, $r_{cav}$ and $r_{duct}$ are the cavity length, the radii of the outer and inner conductor of the cavity, respectively. By transforming the series circuit into RLC parallel circuit (Fig. 4 (b)), the parameters $L$ and $R$ are obtained as

$$2\pi f L = \frac{R_s^2 + (2\pi f L_s)^2}{2\pi f L_s}, \quad (9)$$

$$R = \frac{R_s^2 + (2\pi f L_s)^2}{R_s}. \quad (10)$$

The resonant frequency $f_r$ and Q-value are given by

$$2\pi f_r = \frac{1}{\sqrt{LC}}, \quad (11)$$

$$Q = \frac{1}{2\pi f_r} \frac{R}{L}. \quad (12)$$

The cavity power loss is given by

$$P = \frac{U_g^2}{2R}. \quad (13)$$

The cavity impedance is obtained as

$$Z = \frac{1}{\sqrt{\frac{1}{R^2} + \left( \omega C - \frac{1}{\omega L} \right)^2}}, \quad (14)$$

where angular frequency $\omega = 2\pi f$.

A schematic diagram of the MA-loaded cavity is shown in Fig. 5 and the design specifications are summarized in Table 2. The RF cavity is a single gap structure which consists of two coaxial resonators loaded with MA cores. The length and diameters of inner and outer conductors of the cavity are 550, 150 and 550 mm, respectively. The outer conductor might be fabricated from a metal-mesh plate to cool the MA cores by air cooling. The length of the accelerating gap is 50 mm. The number of MA cores is 8 and the core has an inner diameter of 260 mm, outer diameter of 500 mm and the thickness of 25 mm. The cavity impedance is designed as 490 ± 185 Ω in the operation frequency range of 0.5~7 MHz.

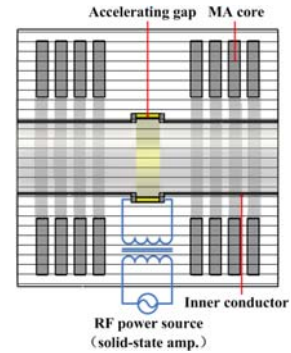

Fig. 5. Schematic diagram of the MA-loaded accelerating cavity.

A solid-state (transistor) amplifier is employed as an RF power source. Impedance matching between the cavity, feeding line and the power source is indispensable to produce a high voltage at the accelerating gap. The impedances of the power source and the feeding line are 50 Ω, and a 1:10 impedance transformer will be used to match the feeding line and cavity.

Table 3 lists the designed parameters for the equivalent circuit (Fig. 4(b)) at the frequencies of 0.5, 1, 3, 5 and 7 MHz. The complex permeabilities of MA cores are the testing data of the 1k107-1 at these frequencies. The shunt impedance $R$ and the inductance $L$ of the cavity are calculated by Eqs. (5)~(10). The equivalent capacitance of the accelerating gap $C$ is calculated by Eq. (4). We determine the size and number of MA cores so that the resonant frequency lies in 0.5~7 MHz. In 1 kV operating voltage, the total power dissipation of MA cores is less than 1.5 kW (<200 W/core). According to Hitachi operating experience [11], forced air-cooling is enough to cool the MA-loaded cavity.

Table 2. Design specifications of the RF cavity

| | |
|---|---|
| Operation frequency | 0.5~7 MHz |
| Gap voltage | 1 kV |
| Cavity impedance | 490 ± 185 Ω (60 ± 25 Ω/core) |
| Cavity structure | Coaxial resonator × 2<br>Cavity length: 550 mm<br>Outer conductor diameter: 550 mm<br>Inner conductor diameter: 150 mm |
| Accelerating gap | Gap length: 50 mm<br>Number of gaps: 1 |
| Core material | 1k107-1 |
| Core shape | Toroidal ring<br>Outer diameter: 500 mm<br>Inner diameter: 260 mm<br>Thickness: 25 mm |
| Number of cores | 8 |
| Cavity power loss | < 1.5 kW<br>< 200 W/core |
| Core cooling | Forced air-cooling |
| Power feeding | 1:10 impedance transformer coupling |

Table 3. Designed parameters of the equivalent circuit representing the RF cavity

| $f$ (MHz) | 0.5 | 1 | 3 | 5 | 7 |
|---|---|---|---|---|---|
| ($\mu_s'$, $\mu_s''$) | (1857, 3159) | (1000, 1931) | (418, 947) | (268, 663) | (200, 514) |
| $R$ (Ω) | 349.75 | 403.13 | 559.88 | 636.91 | 685.54 |
| $L$ (μH) | 188.85 | 123.38 | 66.65 | 49.37 | 39.18 |
| $C$ (pF) | 20 | | | | |
| $Z$ (Ω) | 304.18 | 364.95 | 545.12 | 636.88 | 671.44 |
| $f_r$ (MHz) | 2.59 | 3.20 | 4.36 | 5.06 | 5.68 |
| $P$ (kW) | 1.43 | 1.24 | 0.89 | 0.79 | 0.73 |

### 3.2 Simulation of RF cavity

The CST Microwave Studio is used to simulate the accelerating cavity. Fig. 6 shows the electromagnetic field distribution of accelerating cavity at 5.42 MHz (the complex permeability of MA cores is (268, 663)@5 MHz). As shown in Fig. 6 (a), the electric field is concentrated at the accelerating gap, which satisfies the RF cavity design. From Fig 6 (b), we can see that the magnetic field is distributed almost uniformly in the two co-axial resonators. The main power dissipation of RF cavity is magnetic loss of MA cores.

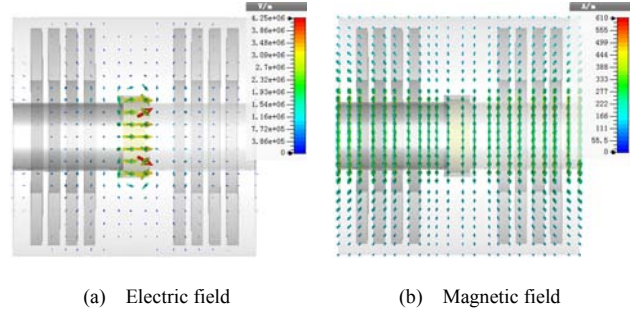

(a) Electric field    (b) Magnetic field

Fig. 6. Electromagnetic field distribution of accelerating cavity.

We use the impedance relative permeability $\mu_z$ [12] for comparing the analytic and CST simulated results

$$\mu_z = \sqrt{\mu_s'^2 + \mu_s''^2}. \tag{15}$$

Fig. 7 shows the $\mu_z$ dependence of analytic and simulated resonant frequencies, and we can see that the two values agree well.

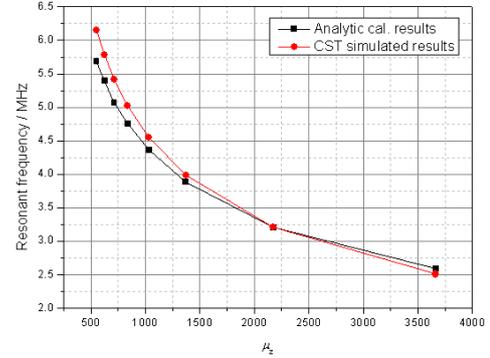

Fig. 7. Analytic and simulated resonant frequencies versus $\mu_z$.

The results of CST simulation are shown in Table 4, in which $f_{cst}$ and $Q_{cst}$, the deviation $f_{rdev}$ between $f_r$ and $f_{cst}$, and $Q_{dev}$ between $Q_{ma}$ and $Q_{cst}$ are given. The difference of $Q$-value is 50~90%.

We also compared the analytic and simulated shunt impedance at resonant frequency $Z_r$ and $Z_{cst}$ in Table 5. $Z_r$ equals $R_r$ at resonant frequency $f_r$, and is rewritten as

$$Z_r = R_r = \frac{R_{sr}^2 + (2\pi f_r L_s)^2}{R_{sr}}, \tag{10'}$$

where $R_{sr}$ is the series resistance at $f_r$

$$R_{sr} = n f_r \mu_0 \mu_s'' t \ln \frac{r_o}{r_i}. \tag{8'}$$

$Z_{cst}$ is estimated at $f_{cst}$ and calculated by

$$Z_{\text{cst}} = \frac{Q_{\text{cst}}}{(2\pi f_{\text{cst}})C}. \quad (16)$$

The deviation $Z_{\text{dev}}$ between $Z_r$ and $Z_{\text{cst}}$ is 50~70%.

Table 4. Analytic and CST simulated results of cavity

| $f$ (MHz) | 0.5 | 1 | 3 | 5 | 7 |
|---|---|---|---|---|---|
| $(\mu_s', \mu_s'')$ | (1857, 3159) | (1000, 1931) | (418, 947) | (268, 663) | (200, 514) |
| $\mu_z$ | 3664 | 2174 | 1035 | 715 | 552 |
| $f_r$ (MHz) | 2.59 | 3.20 | 4.36 | 5.06 | 5.68 |
| $f_{\text{cst}}$ (MHz) | 2.51 | 3.20 | 4.55 | 5.42 | 6.15 |
| $f_{\text{rdev}}$ (%) | -3.17 | 0.01 | 4.50 | 7.12 | 8.19 |
| $Q_{\text{ma}}$ | 0.588 | 0.517 | 0.441 | 0.404 | 0.389 |
| $Q_{\text{cst}}$ | 0.875 | 0.823 | 0.770 | 0.745 | 0.736 |
| $Q_{\text{dev}}$ (%) | 48.7 | 59.0 | 74.6 | 84.5 | 89.0 |

Table 5. Analytic and CST simulated shunt impedance of cavity

| $f_r$ (MHz) | 2.59 | 3.20 | 4.36 | 5.06 | 5.68 |
|---|---|---|---|---|---|
| $Z_r$ (Ω) | 1810 | 1291 | 813 | 645 | 556 |
| $f_{\text{cst}}$ (MHz) | 2.51 | 3.20 | 4.55 | 5.42 | 6.15 |
| $Z_{\text{cst}}$ (Ω) | 2774 | 2043 | 1344 | 1092 | 951 |
| $Z_{\text{dev}}$ (%) | 53.2 | 58.3 | 65.3 | 69.4 | 70.9 |

We think the main reason of these differences is due to the limit of the CST in simulating cavity loaded with high lossy magnetic material. This is the same as that obtained in the previous study [13]. In Ref. 13, it is shown that the analytical results on the cavity impedance agree well with the measurements for the MA-loaded testing cavity where the cavity $Q$-value is ~0.6 [11]. Ref. 13 also shows that the analytical and simulated results on resonant frequencies and $Q$-values agrees very well for $Q > 2$, but not well for $Q < 2$.

## 4 Conclusions

The small-size MA cores made by the Chinese firms have been tested and show good consistency with that of Hitachi material. Based on the MA cores testing results, the RF cavity for the proton and heavy-ion synchrotrons is designed. The frequency range considered is from 0.5 to 7 MHz. The cavity shunt impedance reaches 490 ± 185 Ω, and the total power dissipation with 1kV accelerating voltage is less than 1.5 kW. So forced air-cooling can meet the cooling requirement. Further investigation is required for the Chinese firms to manufacture the large size (outer diameter about 500 mm) cores with similar properties.

The electromagnetic field distributions with CST simulation are very useful for checking the cavity design. The analytic and simulating results on resonant frequencies for MA-loaded accelerating cavity show good agreement. However, for $Q$-factor and cavity impedance, there are some differences between the analytic and simulated results, and it needs the further investigation.